\documentstyle[12pt,psfig,a4,cite]{article}
\begin{document}
\begin{center}
{\bf A Low-Nussinov model for elastic vector meson production at HERA \\
}
~\\
{\bf I. Royen}\\
{\small Inst. de Physique, U. de Li\`ege,
B\^at. B-5, \\
Sart Tilman, B4000 Li\`ege, Belgium\\}
\end{center}

\vskip 0.2 cm
\begin{quote}
I show that a lowest-order pQCD calculation
of elastic vector-meson
production does reproduce all the features of experimental measurements at HERA,
 for all $Q^2$ and all mesons, at fixed energy.
\end{quote}
\vskip 0.2cm

\parskip -0.3 cm
In this model, we work at the level of quarks and gluons and in the high-energy limit, i.e $ w^2 \gg m_V, Q^2, t$.  The amplitude is then greatly simplified to the sum of the discontinuities of two diagrams \cite{JRCIR}.  The model includes three sub-models, for which we first adopted the simplest ones :
we used a non relativistic model for the vector meson, where the quark and the anti-quark share the vector meson momentum ($V$) equally and have a mass equal to $M_V \over {2}$, the colour-singlet
exchange was modeled ``\`a la Low-Nussinov" (two perturbative gluon exchange), and we only
considered the constituent quarks of the proton.  This leads us to two form factors for the proton which cancel the infrared singularity that would result from the pole in the gluon propagator.\\  

We thus obtain an amplitude which is IR finite, conserves s-channel helicity, and is proportional to $w^2$.  This model cannot say anything about the energy dependence of the cross section and we assume that this dependence comes as a fixed factor ($R$) for a given energy.
The longitudinal and transverse part asymptotically behave like $(Q^2)^{-3}$ and $(Q^2)^{-4}$ respectively, although the data collected at HERA are not yet in that asymptotic regime.\\ 

Although this model is very simple, it reproduces the data very well :
 the mass and $Q^2$ dependence of the integrated elastic cross sections, the t-dependence of the $\rho$ photoproduction data, and the ratio of cross sections.  However this model fails, as many existing models, for the ratio $\sigma_L / \sigma_T$ which grows linearly with $Q^2$ whereas the data seem to indicate a plateau around 4 at high $Q^2$.  So the challenge is to get a better prediction for $\sigma_L/\sigma_T$ without losing all the previous results. \\

We decided to push the model further and improve it by introducing a Fermi momentum in the upper loop of the diagrams.  Now the quark and the antiquark don't share the V momentum equally any more and instead of a wave function, we use a vertex function $V_{\mu}\gamma_{\mu}.\phi(l^2)$, with $\phi$ proportional to an exponential :
$\phi(l^2) = N\ exp(l^2/2 p_f^2)$, $2l$ being the 4-momentum describing the relative motion of the quarks.
The proportionality constant is fixed so that we reproduce the leptonic decay rate $\Gamma (V \rightarrow e^{+} e^{-} )$.\\

There is still the question of how we treat the two quarks which form the vector meson.  Of course one of them is still on shell as it is cut when we calculate the discontinuity, and the other one gives two contributions to the amplitude : one from its off-shell behaviour ($A_1$) and one from the discontinuity of its propagator ($A_2$).
The amplitude is then composed of two terms, $A=\sqrt{{A_1}^2+{A_2}^2}$.
The phase is re-absorbed in the normalisation.\\

If we look at the asymptotic forms of the two parts of the amplitude and more precisely at the transverse amplitude, we observe that the second part decreases more slowly with $Q^2$ than before.  
\begin{eqnarray}
A_L^{(1)}&\propto&{128 \pi m_V \over Q^3\ k_t^2}\ \ \ \ \ \ \ \hskip 1.5cm A_T^{(1)}\propto{16 \pi m_q^2 \over Q^4\ k_t^2}\nonumber\\
A_L^{(2)}&\propto&{32 \pi \over Q^3}[a+b\ log(Q^2)]\ \ \ \ \ \ A_T^{(2)}\propto{16 \pi \over Q^4}[c\ Q^2+d\ log(Q^2) +h]
\end{eqnarray}
where $m_q$ is the quark mass.
The total transverse amplitude is then bigger at large $Q^2$ than in the previous model where only the first part of the amplitude was taken into account.  Hence, the ratio $\sigma_L/\sigma_T$ will be smaller.\\

The results that we obtain are as follows : first of all, the dependence on $Q^2$ and $m_V$ of the integrated elastic cross section $\sigma(Q^2)$ [Fig.1 (a)] is such that a common (Regge) factor is consistent with the data
taken at HERA. 
Although we see no reason why our model should work in photoproduction,
it turns out that our curves do go through the measured points.
\\

The ratio of cross sections [Fig.1 (b)] is a more stringent test of
our model, especially as the normalisation then drops out of our prediction.
And, again, we see that our model fares well for $\phi$ over $\rho$, even in photoproduction.\\

The $t$ slope is also well reproduced [Fig.2 (a)].\\

And last but not least, we reproduce the ratio $\sigma_L/\sigma_T$ which doesn't increase with $Q^2$ anymore, but seems now to indicate a plateau for the HERA $Q^2$ range, which is in good agreement with data [Fig.2 (b)].\\

To sum up, this model reproduces : the $Q^2$-dependence of all measured vector meson cross sections,
the mass-dependence of all measured vector meson cross sections,
the ratio $\sigma_L/\sigma_T$, and
the t dependence of $\rho$ photoproduction cross section is also reasonably reproduced.
The novelty is that the interplay between the pole in the quark propagator and the off-shell component needs to be taken into account.

\vskip 0.5 cm
\centerline{
\psfig{figure=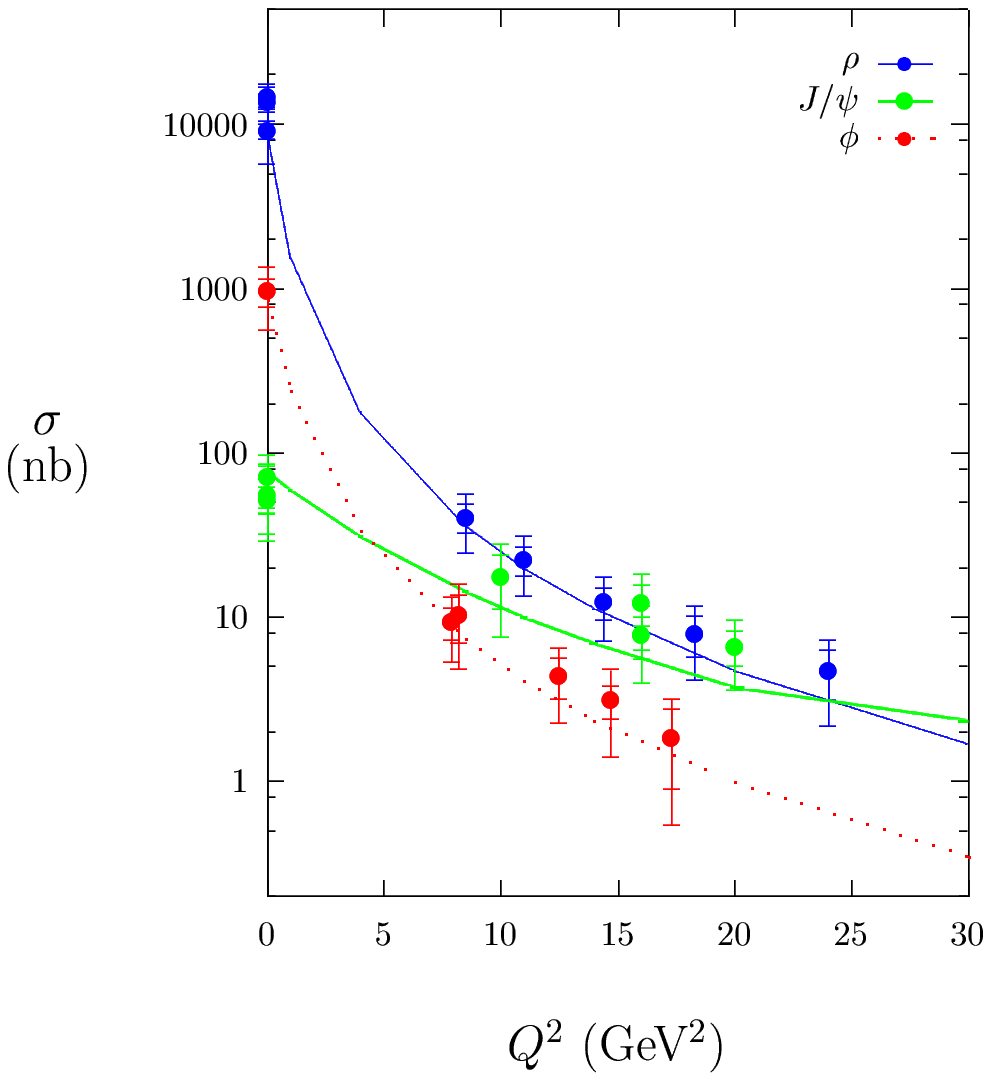,height=5.6cm}\ 
\psfig{figure=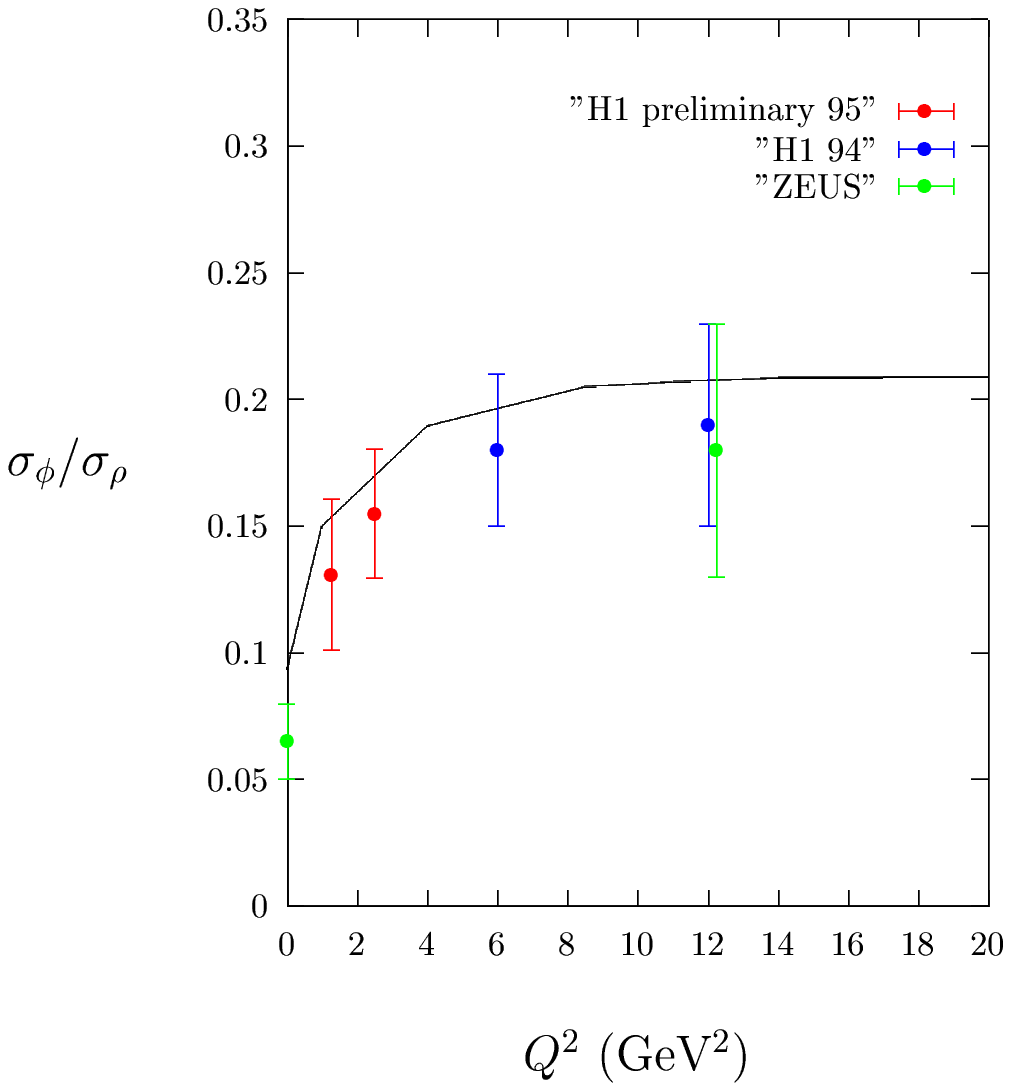,height=5.6cm}}
\begin{quote}
{Figure 1: (a) Cross sections as functions of $Q^2$, (b) Ratio of cross sections as functions of $Q^2$.  Our curves are for $p_f(\rho,\phi,J/\psi)=(0.3,0.3,0.6)$ GeV; $m_q(\rho,\phi,J/\psi)=(0.3,0.45,1.5)$ GeV}
\end{quote}
\vskip 0.5 cm
\centerline{
\psfig{figure=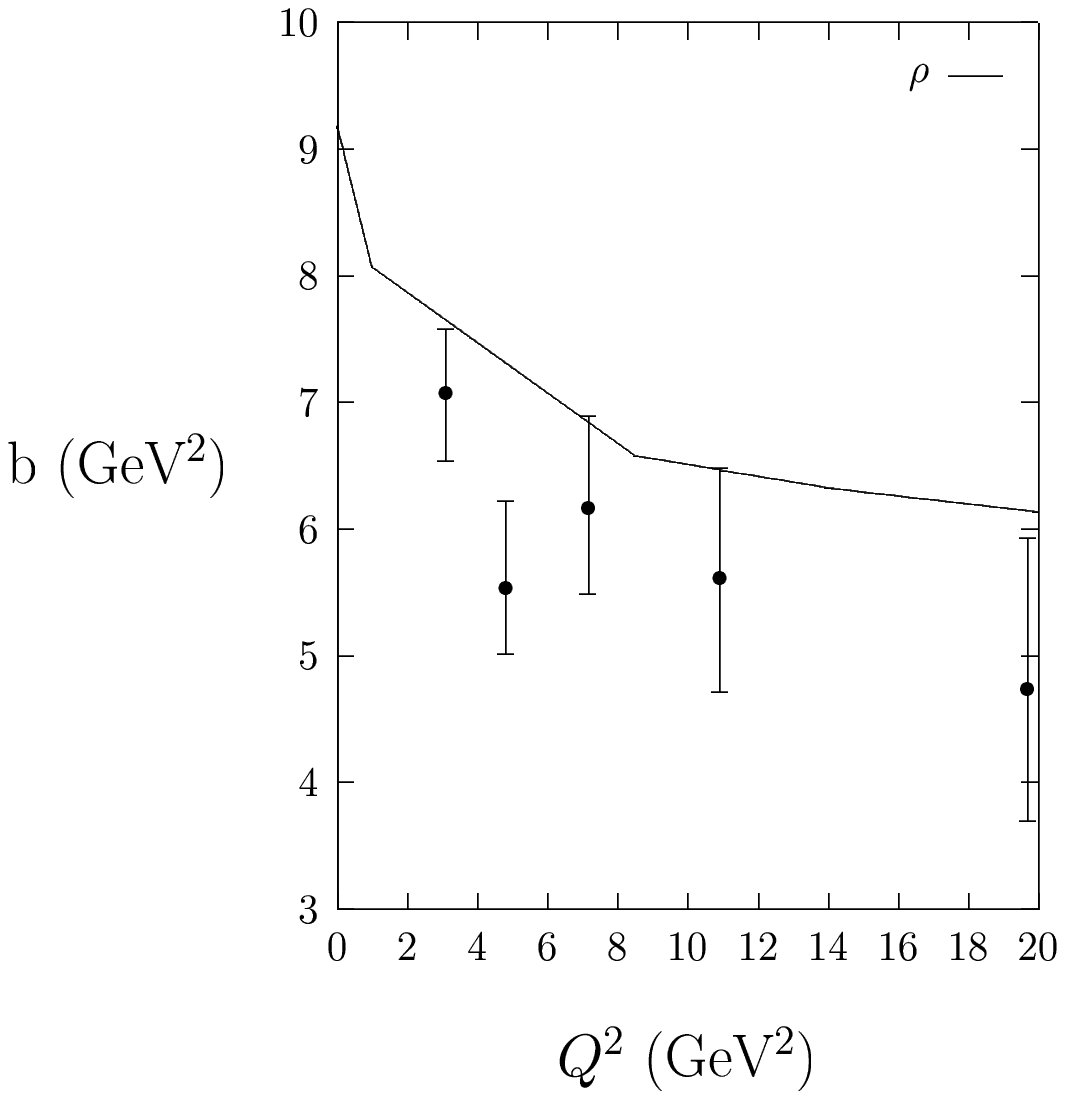,height=5.6cm}\ 
\psfig{figure=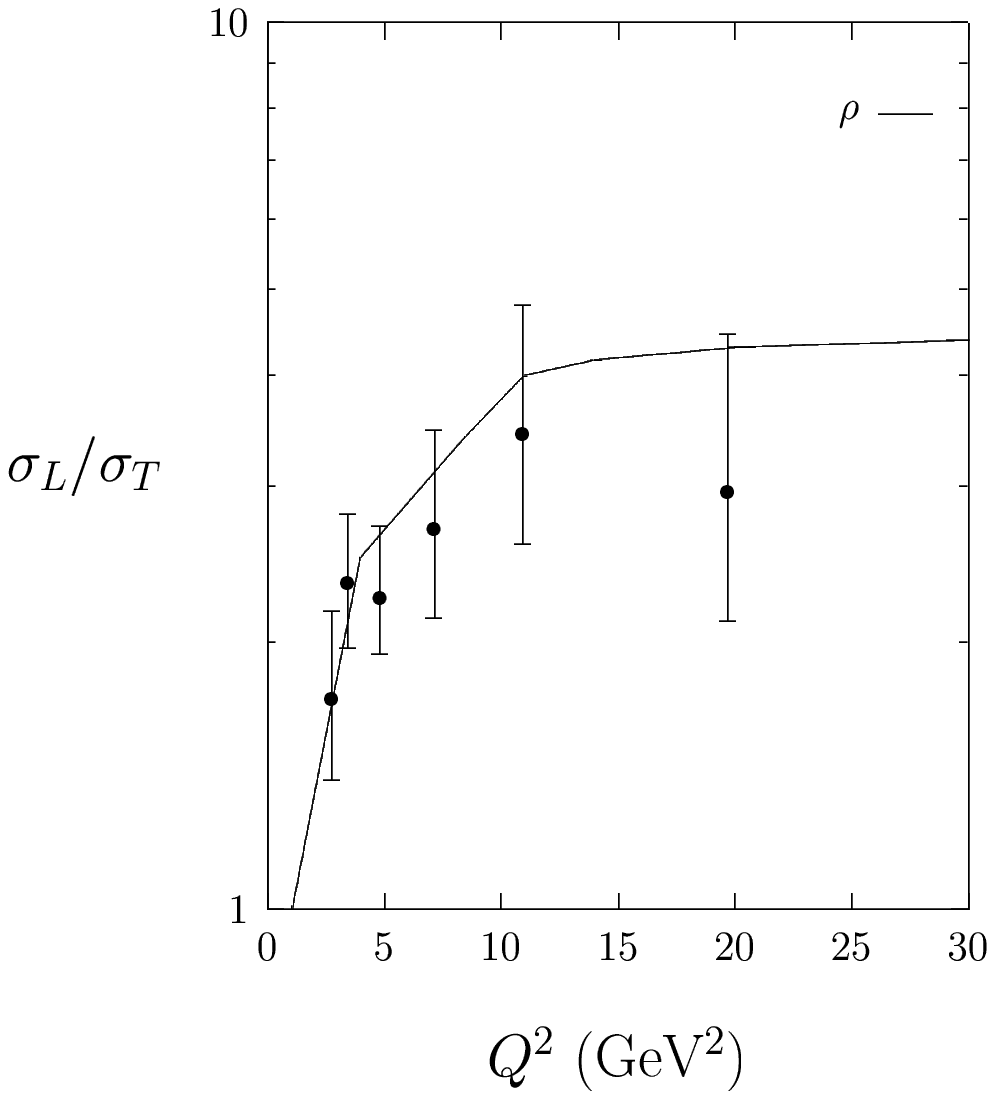,height=5.6cm}}
\begin{quote}
\begin{center}
{Figure 2: (a) $t$-slope for $\rho$, (b) ratio $\sigma_L/\sigma_T$.}
\end{center}
\end{quote}

\end{document}